\documentclass[aps,prb,twocolumn,superscriptaddress,showpacs]{revtex4}
\usepackage{graphics}
\usepackage{subfigure}
\usepackage{longtable}
\usepackage{times}
\usepackage{graphicx}

\usepackage{amsmath}
\usepackage{amsbsy}
\usepackage{dcolumn}

\usepackage{bm}

\usepackage{color}

\begin{document}

\newcommand{\etal}{{\it et al.}\/}
\newcommand{\gtwid}{\mathrel{\raise.3ex\hbox{$>$\kern-.75em\lower1ex\hbox{$\sim$}}}}
\newcommand{\ltwid}{\mathrel{\raise.3ex\hbox{$<$\kern-.75em\lower1ex\hbox{$\sim$}}}}
\def\k{{\bf k}}
\def\rr{{\bf r}}
\def\q{{\bf q}}

\title{Evolution of the neutron resonances in AFe$_2$Se$_2$}

\author{T.~A.~Maier} \affiliation{Center for Nanophase Materials Sciences and Computer Science and Mathematics Division, Oak Ridge National Laboratory, Oak Ridge, TN 37831-6494}
\author{P.~J.~Hirschfeld} \affiliation{Department of Physics, University of Florida, Gainesville, FL 32611, USA}
\author{D.~J.~Scalapino} \affiliation{Department of Physics, University of California, Santa Barbara, CA 93106-9530 USA}

\date{\today}
\begin{abstract}
	Recent experiments on the alkali-intercalated iron selenides have raised questions about the symmetry of the superconducting phase. Random phase approximation calculations of the leading pairing eigenstate for a tight-binding 5-orbital Hubbard-Hund model of AFe$_2$Se$_2$ find that a $d$-wave ($B_{1g}$) state evolves into an extended $s^\pm$ ($A_{1g}$) state as the system is hole-doped. However, over a range of doping these two states are nearly degenerate. Here, we calculate the imaginary part of the magnetic spin susceptibility $\chi''(q,\omega)$ for these gaps and discuss how the evolution of neutron scattering resonances can distinguish between them.
\end{abstract}

\pacs{74.70.-b,74.25.Ha,74.25.Jb,74.25.Kc}

\maketitle

\section{Introduction\label{sec:1}}

Initial experimental reports on the alkali-intercalated iron selenide materials of nominal composition AFe$_2$Se$_2$\cite{firstreports} indicated a number of surprising results\cite{trend} which apparently differentiated these materials from their iron chalcogenide and iron pnictide superconductor cousins. In contrast to these last systems, nearly all of which have hole and electron Fermi surface pockets present simultaneously, angle-resolved photoemission (ARPES) studies suggested that there were no hole-like Fermi sheets for some dopings.\cite{HongDing} Secondly, the resistivity of these samples were all remarkably large, suggesting a proximity to a metal-insulator transition. Finally, a phase of ordered Fe vacancies supporting a large antiferromagnetic ordered moment arranged in a $\sqrt5\times\sqrt5$ block spin pattern was reported by elastic neutron scattering.\cite{WBao} The implications of all these anomalous features for superconductivity in these systems, which occurs at temperatures as high as 31K, is not presently clear. Indeed, it is not even established whether superconductivity coexists with the vacancy ordered phase, arises in Fe vacancy-free regions, or is found in disordered vacancy regions of the multiphase samples.

From the theoretical standpoint, various authors constructed models based on the early ARPES data and investigated new physics associated with the absence of the hole pockets. Within spin fluctuation theory, the consequences of this Fermi surface topology was already discussed by Kuroki \etal,\cite{k_kuroki_08} who pointed out that this band structure is a 2D version of ``Agterberg-Barzykin-Gor'kov" nodeless $d$-wave superconductivity \cite{ABG}, since the symmetry-enforced nodal lines fall between the Fermi surfaces. This idea was adopted in the context of the new systems by several authors. Wang \etal\cite{Lee} predicted based on a functional renormalization group (fRG) calculation that a $d$-wave state would be favorable in such a situation, with an $s^\pm$ state a close competitor, and suggested that the latter possibility was due to the ``marginal" hole band just below the Fermi level in their calculations. Maier \etal\cite{GraserKFeSe} performed similar calculations in the random phase approximation (RPA) and also found competing $d-$ and $s-$wave order for large electron doping, but only $d$-wave pairing stabilized in the absence of hole Fermi pockets\cite{Maiti}. These authors also found a strong peak in the dynamical susceptibility not at the wave vector ${\bf Q}=(\pi,\pi)$ corresponding to the nesting wave vector of the two electron pockets in the 1-Fe Brillouin zone, but rather close to $(\pi,0.6\pi)$, the vector connecting the closest flat sides of the rather square electron pockets. Mazin \cite{Mazin2011} and Khodas and Chubukov \cite{Khodas2012} have discussed the role of hybridization and the appearance of an $s^\pm$ state, in which the gap changes sign between the hybridized electron pockets. Other authors considered weak-coupling models involving proximity to or coexistence with simple magnetic stripes,\cite{DasBalatsky}  with the $\sqrt{5}\times \sqrt{5}$ block state,\cite{blockAF_SC} or considered similar electronic structure in orbital fluctuation pairing models\cite{Saito}. Strong-coupling models predicting fully gapped $s$-like states have also been proposed\cite{Yuetal,Fangetal}.

A final piece of important experimental information was obtained when inelastic neutron scattering measurements on Rb$_x$Fe$_{2-y}$Se$_2$ performed by Park \etal\cite{Inosov1} reported a resonance in the superconducting state similar to that observed in other Fe-pnictide and Fe-chalcogenide superconductors\cite{neutronreview} {\it except} that it was observed not at ${\bf Q}=(\pi,0)$ (in the 1-Fe Brillouin zone) but at ${\bf Q}\simeq (\pi, \pi/2)$, very close to the value predicted by Maier \etal\cite{GraserKFeSe} Friemel \etal\cite{Inosov2} then observed a dispersion of the resonant mode consistent with band structure calculations on RbFe$_2$Se$_2$, within the RPA $d$-wave spin fluctuation picture. This represents strong support for the itinerant nature of the mode, and its ultimate origin in the strong scattering between the electron Fermi pockets in the doped metallic phase. In these experiments, there was no sign of the iron-vacancy ordering; these authors concluded that their signal was coming from nonmagnetic, vacancy-disordered or free phases, and that an itinerant picture of the phenomenon was essentially correct.

It is important to establish experimentally whether or not a $d$-wave state is realized in this system, since there is accumulating evidence that all other Fe-based superconductors have $s$-wave (albeit probably $s_\pm$) symmetry.  Observation of a different symmetry in this case could be consistent with the predictions of spin-fluctuation theory, which finds an increasingly competitive $d$-wave instability as hole pockets shrink and disappear\cite{Maiti}, or it could represent a genuinely new paradigm related to the other unusual features of the alkali-intercalated chalcogenides.  Recently, Xu et al.\cite{DLFeng} reported an ARPES measurement on a KFe$_{2-y}$Se$_2$ sample with a small $Z$-centered hole pocket.  If a $d$-wave gap is present, it must by symmetry possess nodes on such a pocket.  However, the authors of Ref. \onlinecite{DLFeng} were unable to detect significant gap anisotropy on this pocket, and concluded that the data were inconsistent with $d$-wave gap symmetry. On the other hand, synchrotron-based ARPES has had a great deal of difficulty observing gap anisotropy in these systems, even in situations where other probes have provided strong evidence for gap nodes (for a discussion, see Ref. \onlinecite{HKM}).  In addition, we argue below that it is difficult to reconcile the inelastic neutron scattering data with an $s$-wave state.  We therefore regard the question of the symmetry of the gap in these systems as open.

This paper is concerned with exploring ways in which further neutron scattering experiments on similar systems could provide information on the gap structure. We are particularly interested in how such experiments can distinguish $A_{1g}$ $s$-wave and $B_{1g}$ ($d$-wave) gap structures in systems where the underlying electronic structure consists of two electron sheets and small hole pockets which could arise upon hole doping. When the AFe$_2$Se$_2$ system is doped so as to move the hole bands through the Fermi surface, one expects that as the conventional Fe-pnictide type Fermi surface, with two electron and two hole pockets is recovered, the system will make a transition from a $B_{1g}$ ($d$-wave) state to an $A_{1g}$ ($s^\pm$-wave) state. However, RPA calculations find that the $B_{1g}$ pairing is surprisingly robust and that a near-degeneracy between the $A_{1g}$ and $B_{1g}$ channels occurs over a finite doping range where the hole pockets first appear. It is in this doping range that we will examine how neutron scattering can provide information on the gap structure.

\section{Model}

\begin{figure}
	[tbp]
	\includegraphics[width=0.75\columnwidth]{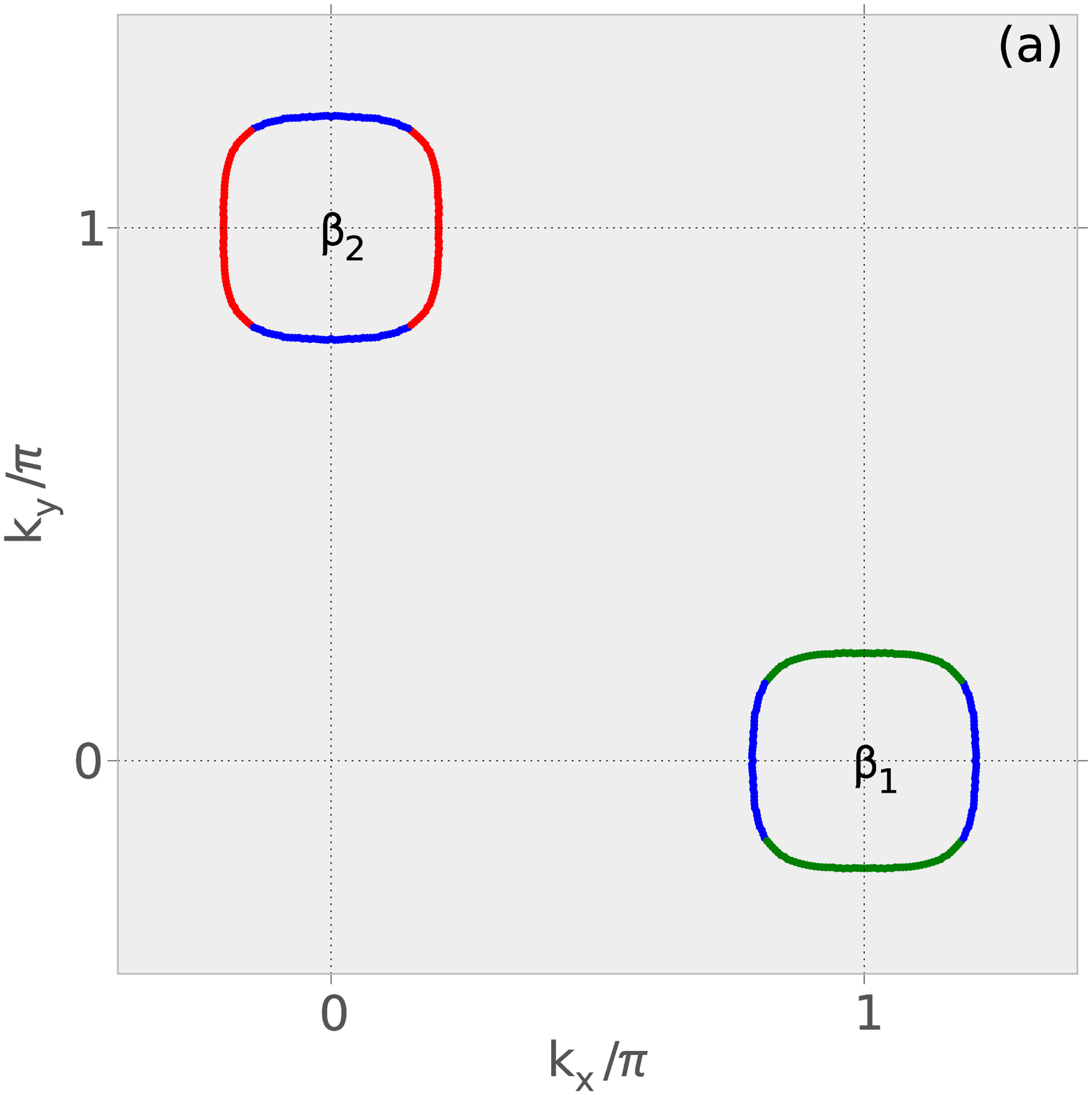}
	\includegraphics[width=0.75\columnwidth]{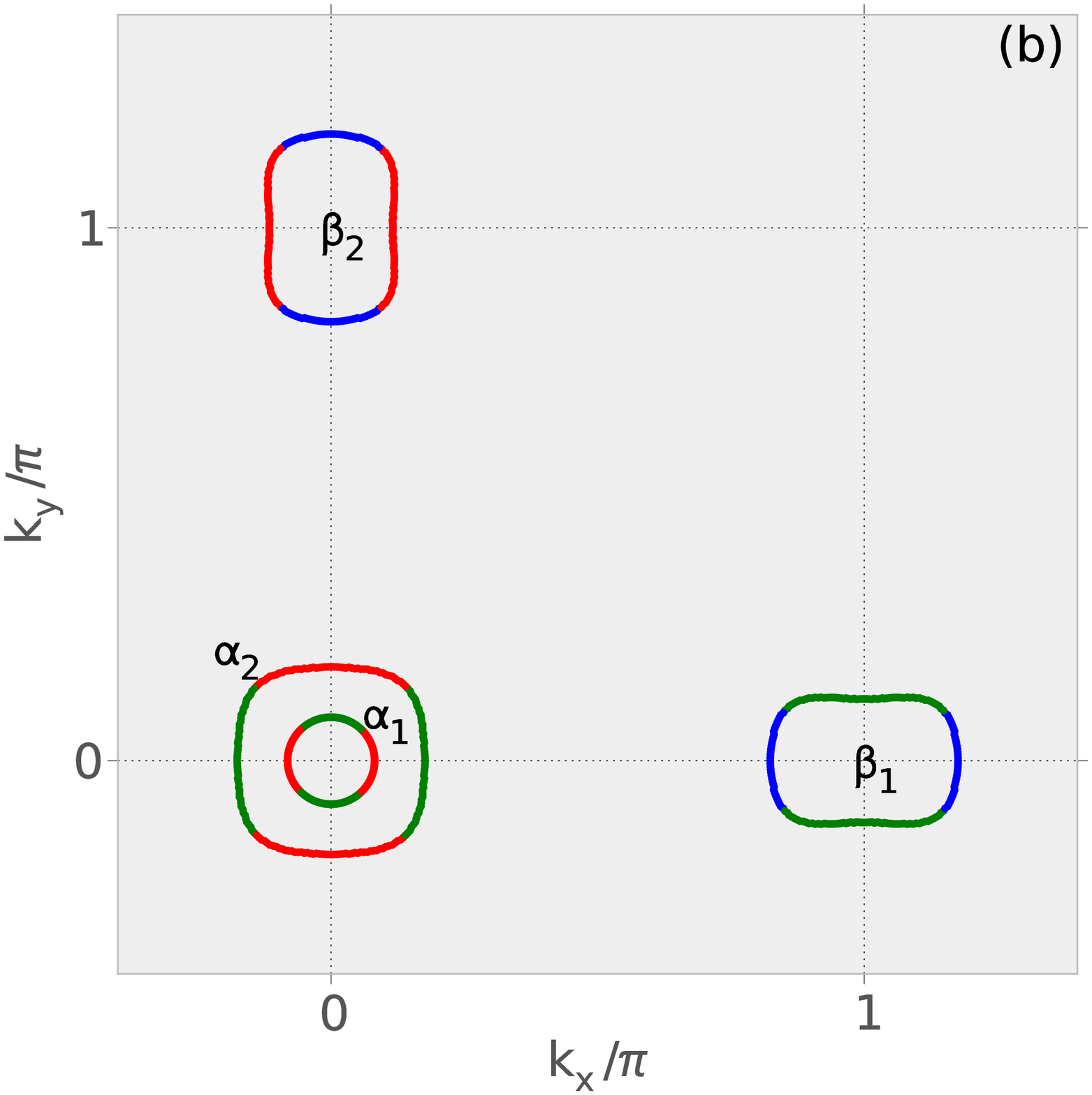} \caption{(Color online) Fermi surfaces and orbital weights for (a) $\langle n\rangle=0.15$ and (b) $\langle n\rangle=0$. In the latter case the filling is such that hole pockets appear around the $\Gamma$ point. Colors represent dominant orbital content: red (xz), green (yz) and blue (xy). \label{fig:scattering}}
\end{figure}

In the following, we consider the 5-orbital Hubbard-Hund Hamiltonian
\begin{eqnarray}
	H = H_{0}& + &\bar{U}\sum_{i,\ell}n_{i\ell\uparrow}n_{i\ell\downarrow}+\bar{U}'\sum_{i,\ell'<\ell}n_{i\ell}n_{i\ell'} \nonumber\\
	& + & \bar{J}\sum_{i,\ell'<\ell}\sum_{\sigma,\sigma'}c_{i\ell\sigma}^{\dagger}c_{i\ell'\sigma'}^{\dagger}c_{i\ell\sigma'}c_{i\ell'\sigma}\\
	& + & \bar{J}'\sum_{i,\ell'\neq\ell}c_{i\ell\uparrow}^{\dagger}c_{i\ell\downarrow}^{\dagger}c_{i\ell'\downarrow}c_{i\ell'\uparrow} \nonumber \label{H}
\end{eqnarray}
where the interaction parameters $\bar{U}$, $\bar{U}'$, $\bar{J}$, $\bar{J}'$ are given in the notation of Kuroki \etal \cite{Kuroki08}. Here we have used spin rotational invariant parameters ${\bar U}=0.92$, ${\bar U}'={\bar U}/2$ and ${\bar J}={\bar J}'={\bar U}/4$. The tight-binding Hamiltonian $H_0$ was fitted to the full DFT band structure of the parent compound KFe$_2$Se$_2$, and the splitting between the two $d_{xz}/d_{yz}$ bands and the two $d_{xy}$ bands at the $\Gamma$ point was artificially enhanced to account for the ARPES results \cite{ref:Qian} as described in Ref.~\onlinecite{GraserKFeSe}. The Fermi surfaces and orbital weights $\left|\langle d_\ell|\nu k\rangle\right|^2$ for electron dopings of $n=0.15$ and 0.05 are shown in Fig.~\ref{fig:scattering}. Here, $\ell$ is an orbital index with $\ell\in(1,2,3,4,5)$ corresponding to the Fe-orbitals $(d_{xz},d_{yz},d_{xy},d_{x^2-y^2},d_{3z^2-r^2})$ and $\nu$ and $k$ denote the band and wave vector of the Bloch states. The pair scattering processes between the $\beta$ Fermi surfaces of Fig.~\ref{fig:scattering}a promote $B_{1g}(d_{x^2-y^2})$ pairing while scattering processes between the $\alpha$ and $\beta$ Fermi surfaces of Fig.~\ref{fig:scattering}b give rise to $A_{1g}(s^\pm)$ pairing. Note that the square Fermi surface pockets found here allow for the possibility of nesting at vectors away from $(\pi,\pi)$ between the two electron $\beta-$ pockets, but that it is the contribution to the real part of the bare susceptibility $\chi_0$ due to the opposite signs of
the Fermi velocities on these two nearly parallel Fermi surface edges which provide the dominant enhancement of
$\Gamma_{ij}$ within the RPA.

Within the spin fluctuation exchange approach, the pairing symmetry for a given set of parameters is determined by the eigenvector of the leading eigenvalue of the scattering vertex $\Gamma(\k,\k')$ in the singlet channel,
\begin{eqnarray}
	{\Gamma}_{ij} (\k,\k') & = & \mathrm{Re}\sum_{\ell_1\ell_2\ell_3\ell_4} a_{\nu_i}^{\ell_2,*}(\k) a_{\nu_i}^{\ell_3,*}(-\k) \\
	&&\times \left[{\Gamma}_{\ell_1\ell_2\ell_3\ell_4} (\k,\k',\omega=0) \right] a_{\nu_j}^{\ell_1}(\k') a_{\nu_j}^{\ell_4}(-\k').\nonumber \label{eq:Gam_ij}
\end{eqnarray}
Here $\k$ and $\k'$ are momenta restricted to the electron and hole pockets $\k \in C_i$ and $\k' \in C_j$, where $i$ and $j$ correspond to either the $\alpha$ or $\beta$ Fermi surfaces, and $a^{\ell}_{\nu}(\k)=\langle d_\ell|\nu k\rangle$ are orbital-band matrix-elements. The vertex function in orbital space $\Gamma_{\ell_1\ell_2\ell_3\ell_4}$ represent the particle-particle scattering of electrons in orbitals $\ell_1,\ell_4$ into $\ell_2,\ell_3$ and in an RPA approximation is given by:
\begin{eqnarray}
	&&{\Gamma}_{\ell_1\ell_2\ell_3\ell_4} (\k,\k',\omega) = \left[\frac{3}{2} \bar U^s \chi_1^{\rm RPA} (\k-\k',\omega) \bar U^s + \nonumber \right.\,~~~~~~\,\\
	&&\,~~~~~\left. \frac{1}{2} \bar U^s - \frac{1}{2}\bar U^c \chi_0^{\rm RPA} (\k-\k',\omega) \bar U^c + \frac{1}{2} \bar U^c \right]_{\ell_1\ell_2\ell_3\ell_4}. \label{eq:fullGamma}
\end{eqnarray}
The interaction matrices $\bar U^s$ and $\bar U^c$ in orbital space are constructed from linear combinations of the interaction parameters, and their forms are given e.g.\/ in Ref.~\onlinecite{ref:Kemper}. Here, $\chi_1^{\rm RPA}$ and $\chi_0^{\rm RPA}$ are the spin-fluctuation and the orbital-fluctuation parts of the RPA susceptibility, respectively.

Then, the pairing strength\cite{Graser09} $\lambda_\alpha$ for various pairing channels $\alpha$ are given as eigenvalues of

\begin{equation}
	\label{eqn:gapeqn} - \sum_j \oint_{C_j} \frac{d \k_\parallel'}{2\pi v_F(\k_\parallel')} \Gamma_{ij} (\k,\k') g_\alpha (\k') = \lambda_\alpha g_\alpha(\k).
\end{equation}
The eigenfunction $g_\alpha(\k)$ for the largest eigenvalue determines the leading pairing instability and provides an approximate form for the superconducting gap $\Delta(\k)\sim g(\k)$.

At the doping $\langle n\rangle =0.15$ shown in Fig.~\ref{fig:scattering}a, the $\alpha$ hole pockets surrounding the $\Gamma$ point are absent, and the leading pairing instability of Eq.~(\ref{eqn:gapeqn}) occurs in the $B_{1g}$ ($d_{x^2-y^2}$) channel. However, as holes are added and $\langle n\rangle$ decreases, the $\alpha$ hole pockets appear as shown in Fig.~\ref{fig:scattering}b for $\langle n\rangle=0$, and the leading pairing instability can occur in the $A_{1g}$ ($s^\pm$) channel.

\begin{figure*}
	[tbp]

	\includegraphics[height=8in]{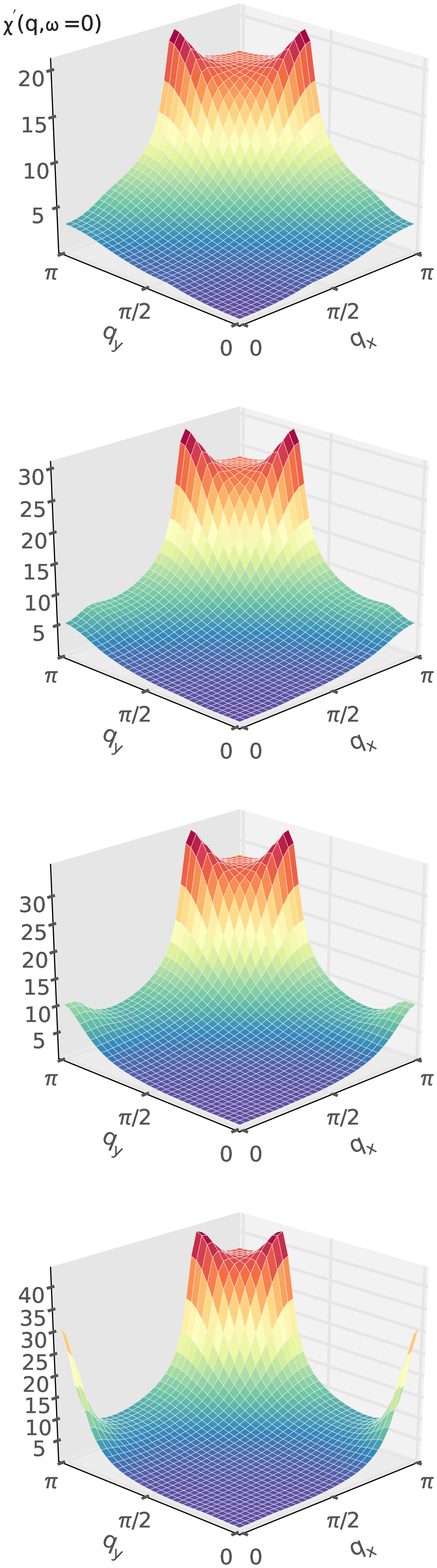}
	\includegraphics[height=8in]{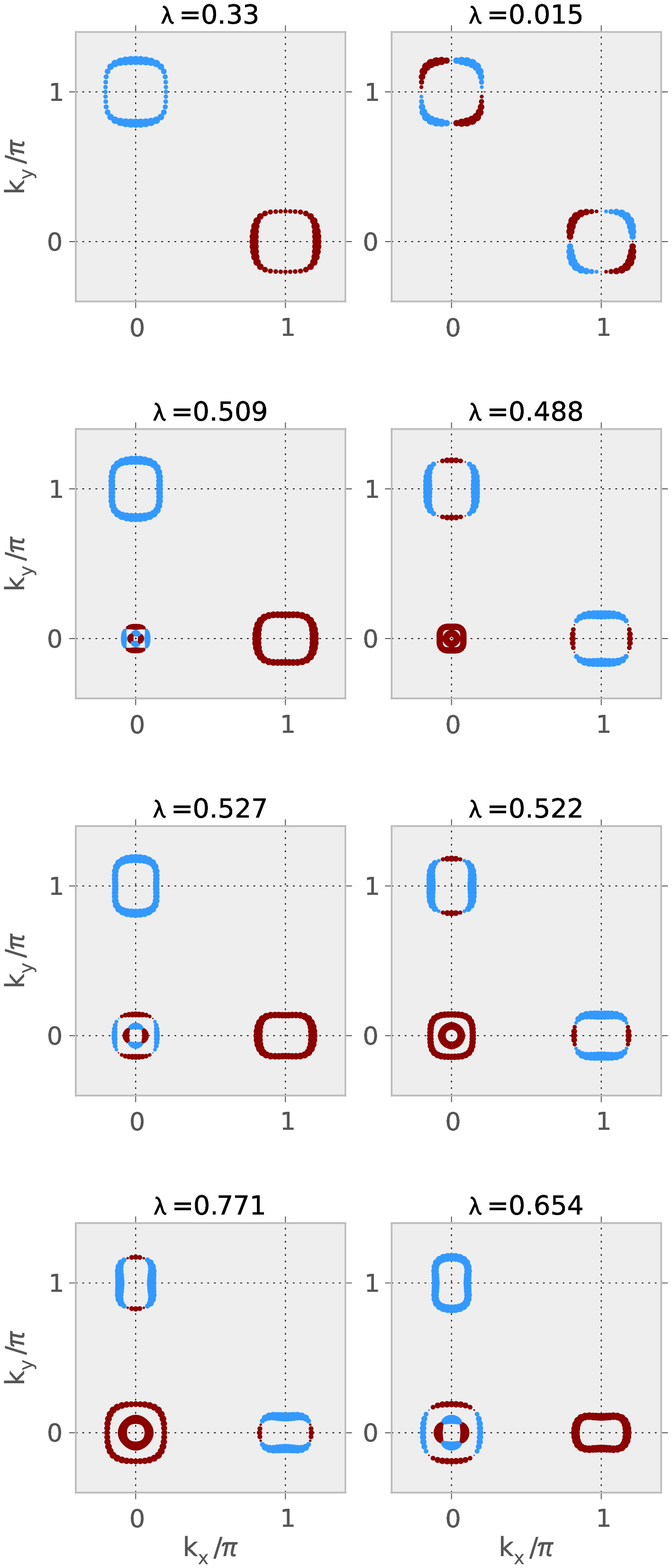} \caption{(Color online) Left column: $\chi({\bf q},\omega=0)$ for (top to bottom) dopings $n=$ 0.15, 0.09, 0.05, 0.0 electrons. Next two columns: Leading and next leading eigenvector $g_{\bf k}$ at the same dopings. Red and blue colors indicate positive and negative values, while the symbol size reflects the magnitude. For the interaction parameters that we have taken, the $d$-wave and extended $s^\pm$ states are nearly degenerate for a range of dopings with a change from $d$ to $s^\pm$ pairing occuring between $n=0.05$ and $0.0$. \label{fig:fig2}}
\end{figure*}

Figure~\ref{fig:fig2} illustrates the momentum dependence of the RPA spin susceptibility which drives the pairing along with the momentum space structure of the two leading eigenfunctions on the Fermi surfaces for four dopings. Here one sees that over a range of dopings from 0.1 to 0.0, these two eigenvalues are remarkably close.
Thus with doping or possibly pressure there can be an evolution from a $B_{1g}$ ($d_{x^2-y^2}$) to an $A_{1g}$ ($s_{\pm}$) state with even the possibility of a $d+is$ state \cite{Zhang,Hanke}.

As such an evolution proceeds, one would expect to find a change in structure of the neutron scattering resonant response. In the following we calculate the susceptibility in the symmetry-broken state as~\cite{ref:Maieretal}
\begin{eqnarray}
	\chi^0_{\ell_1\ell_2\ell_3\ell_4} (q) & = & -\frac{T}{2N} \sum_{k,\mu\nu} M^{\mu\nu}_{\ell_1\ell_2\ell_3\ell_4} (\k,\q) \\
	& & \times \{ G^{\mu}(k+q) G^{\nu} (k) + F^{\mu}(-k-q) F^{\nu} (k) \} \nonumber \label{eqn:supersuscept}
\end{eqnarray}
where the 4-momenta are $q=(\q,\omega_m)$ and $k=(\k,\omega_n)$. The normal and anomalous Green's functions are given as
\begin{equation}
	G^\mu(k)= \frac{i \omega_n + \xi_\nu(\k)}{\omega_n^2 + E_\nu^2(\k)}, \;\;\; F^\mu(k)= \frac{\Delta(\k)}{\omega_n^2 + E_\nu^2(\k)}
\end{equation}
Here the matrix elements relating band and orbital space are
\begin{equation}
	M^{\mu\nu}_{\ell_1\ell_2\ell_3\ell_4} (\k,\q) = a_\mu^{\ell_1} (\k+\q) {a_\nu^{\ell_2}}^* (\k) {a_\mu^{\ell_3}}^* (\k+\q)a_\nu^{\ell_4} (\k) 
\end{equation}
and the quasiparticle energies for band $\nu$ are given by $E_\nu(\k) = \sqrt{\xi_\nu^2(\k) + \Delta^2(\k)}$. The neutron scattering cross section is proportional to the imaginary part of the spin susceptibility in the superconducting state
\begin{equation}
	\label{eqn:chisum} \chi (\q,\omega) = \sum_{\ell_1 \ell_2} \chi^{\rm RPA}_{\ell_1\ell_1\ell_2\ell_2} (\q,\omega)\,.
\end{equation}
The multiorbital RPA spin susceptibility is now given by
\begin{equation}
	\label{eqn:RPA} \chi^{\rm RPA}_{\ell_1\ell_2\ell_3\ell_4} (q,\omega) = \left\{ \chi^0 (q,\omega) \left[1 -\bar U^s \chi^0 (q,\omega) \right]^{-1} \right\}_{\ell_1\ell_2\ell_3\ell_4} \,.
\end{equation}

To understand the type of structure that one can expect to see in inelastic neutron scattering experiments for such a system in the $d_{x^2-y^2}$, $s^\pm$ and possible intermediate $d+is$ states, we have calculated Im $\chi(\q,\omega)$ from Eqs.~(\ref{eqn:chisum}-\ref{eqn:RPA}) for some simple parametrizations of $\Delta(\k)$.
As seen in Fig.~\ref{fig:fig2}, the $A_{1g}$ gap exhibits nodes on the electron sheets and has a gap which we will parameterize using an extended $s$-wave ($xs$) $(\cos k_x+\cos k_y)$ form. We will also consider a simpler isotropic $s^\pm$ gap. The $d$-wave gap has the expected $(\cos k_x-\cos k_y)$ form except that the gap on the $\alpha_2$ Fermi surface has a phase factor of $-1$. This is because the spin fluctuation scattering of $d_{yz}$ pairs from both $\alpha_1$ and $\alpha_2$  to the $\beta_2$ Fermi surface  provides the dominant contribution to the interaction, along with  
the scattering of $d_{xz}$ pairs  to the $\beta_1$ Fermi surface. For a $d_{x^2-y^2}$ gap this means that the sign of the gaps on the regions of the $\alpha_1$ and $\alpha_2$ Fermi surfaces where the $d_{yz}$ orbital weight is largest will be opposite to that where the $d_{xz}$ orbital weight is largest. Thus the anti-phase appearance of the $d$-wave gaps on the $\alpha_2$ hole Fermi surface is simply a reflection of the $d_{xz}$ and $d_{yz}$ orbital weights, which as seen in Fig.~\ref{fig:scattering} are out of phase by $\pi/2$ relative with these weights on the $\alpha_1$ Fermi surface.

With these considerations in mind, we have parameterized the gaps as follows:

\begin{eqnarray}
	\Delta^{s^\pm}_\nu &=& \Delta_\nu \label{eqn:gaps1}\\
	\Delta^{xs}_\nu(\k) &=& \Delta_\nu(\cos k_x + \cos k_y)\label{eqn:gaps2} \\
	\Delta^d_\nu(\k) &=& \Delta_\nu (\cos k_x - \cos k_y)\label{eqn:gaps3} \\
	\Delta^{d+is}_\nu(\k) &=& \left(\Delta^d_\nu(\k)+i \Delta^{xs}_\nu(\k)\right)/\sqrt{2}\label{eqn:gaps4}
\end{eqnarray}
Here the Fermi surface sheet dependent gap amplitudes $\Delta_\nu$ are adjusted so that the maximum amplitude is 0.05 on each sheet. For the $s^\pm$ gap, $\Delta_\nu$ on the $\alpha$ sheets is positive and on the $\beta$ sheets negative. For the $d$-wave, $\Delta_{\alpha_2}$ is negative (out of phase) with respect to the $\Delta_{\alpha_1}$ gap. In the $\k$-space integrations, Eq.~\ref{eqn:supersuscept}, the gaps are cut off using $\exp\left(\left(\xi_\nu(\k)-\mu\right)^2/\Omega^2_0\right)$ as $\k$ moves away from the Fermi surface with $\Omega_0=0.1$ eV.

It has been argued that the $d$-wave $\cos k_x - \cos k_y$ for the gap on the $\beta$ Fermi surfaces is fragile and acquires nodes once the Se mediated hybridization between the electron pockets is taken into  account\cite{Mazin2011,Khodas2012}. However the nodal regions do not make a significant contribution to the spin resonance and the simple $d$-wave form of  Eq.~(\ref{eqn:gaps3}) provides a suitable representation of the B1g gap for the 5-orbital one Fe per unit cell model. The hybridization can also  alter the usual $A_{1g}$ $s^\pm$ gap giving rise to an $A_{1g}$ gap which changes signs between the hybridized $\beta$ pockets. The neutron scattering from this state is expected to be similar to that of the extended $s$-wave case, Eq.~(\ref{eqn:gaps2}).

\begin{figure}
	[h]
	\includegraphics[width=0.95\columnwidth]{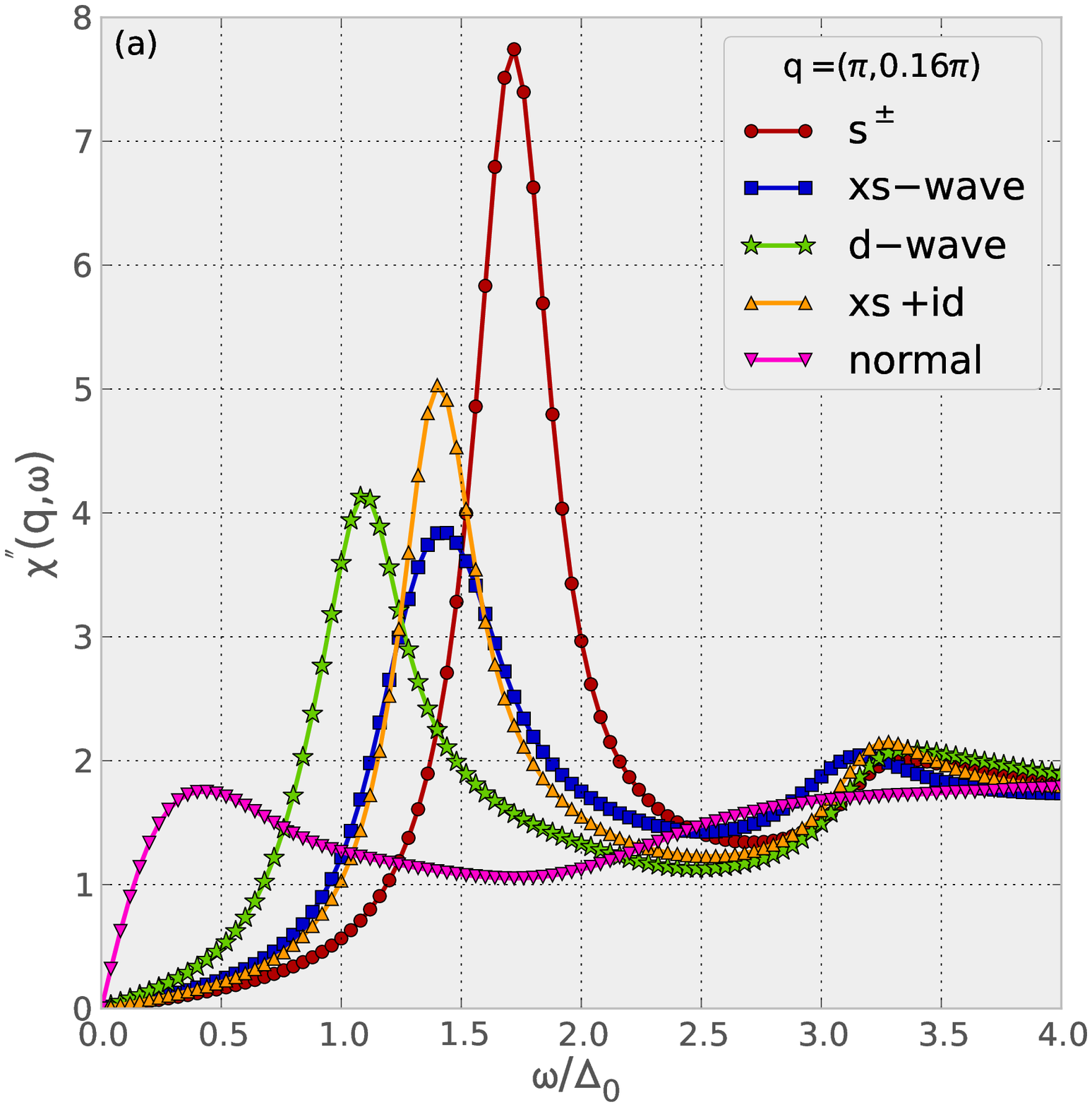}
	\includegraphics[width=0.95\columnwidth]{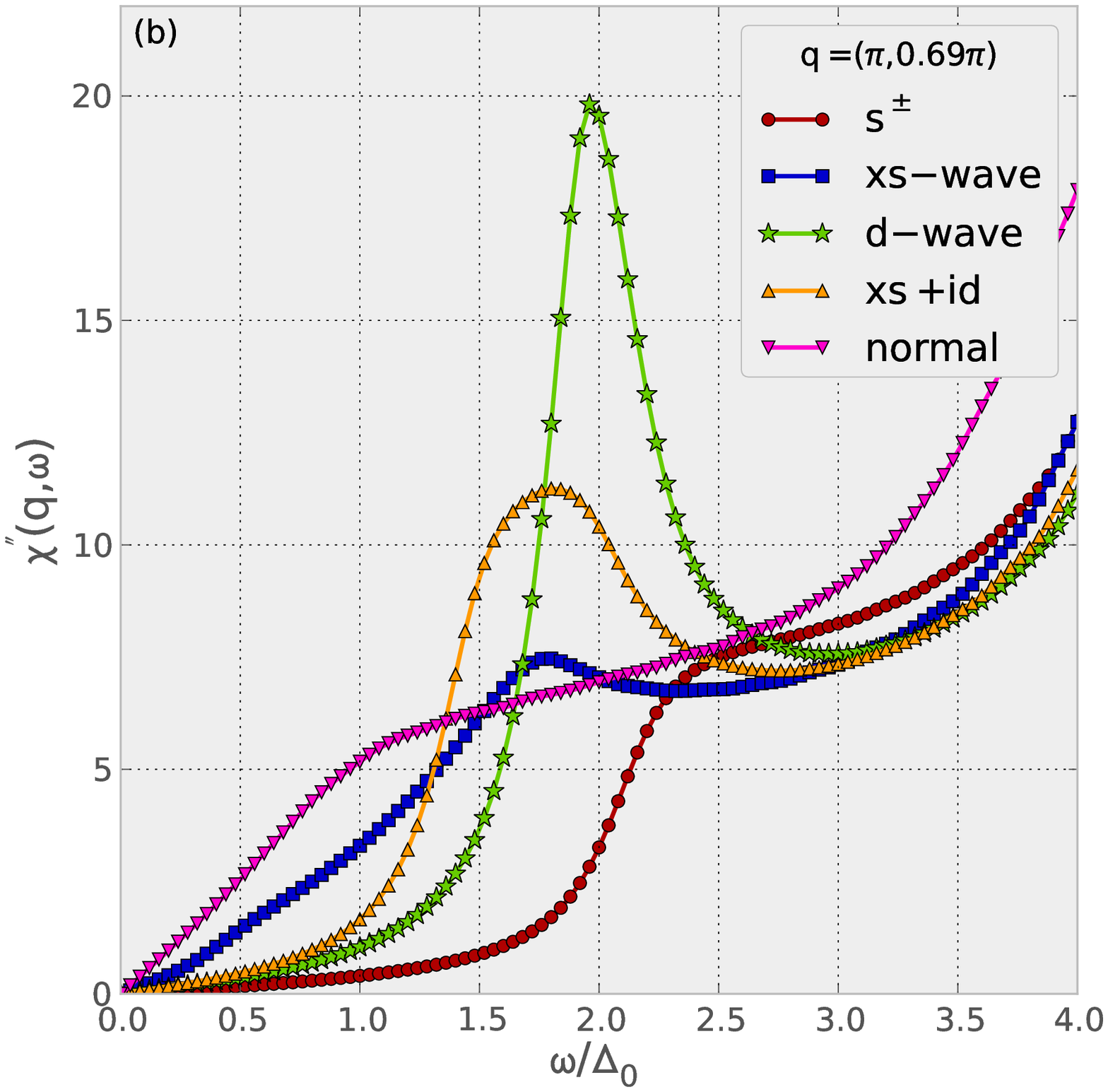} \caption{(Color online) Comparison of neutron response of superconducting states given in Eq.~(\ref{eqn:gaps1})--\ref{eqn:gaps4}) for single doping $\langle n \rangle=0.09$ ($\mu=-0.0305$). Left: $\chi''({\q=(\pi,0.16\pi)},\omega)$; Right: $\chi''({\q=(\pi,0.69\pi)},\omega)$. \label{fig:fig3}}
\end{figure}

To illustrate how the gap structure is reflected in the inelastic neutron scattering, in Fig.~\ref{fig:fig3} we show the neutron scattering response $\chi''(q,\omega)$ for an electron doping $n=0.09$. This is a doping in the range where the $A_{1g}$ and $B_{1g}$ instabilities are close to each other and where small changes can lead to an evolution of one state into another or to even the formation of an $xs+id$ state. Thus it represents an interesting testing ground to explore the neutron scattering resonances. Results for momentum transfers $q_1=(\pi,0.16\pi)$ and $q_2=(\pi,0.7\pi)$, corresponding to the 
peaks in the dynamical sysceptibility of Fig. \ref{fig:fig2},
are shown in Figs.~\ref{fig:fig3}a and b, respectively. The $s^\pm$ gapped phase shows the expected resonance response for a momentum transfer $q_1$ in which an electron is scattered from the $\alpha_2$ Fermi surface where the gap is positive to the $\beta_1$ Fermi surface where the gap is negative. The extended $s$-wave $xs$ gap shows a similar resonance which arises from scatterings between the $\alpha_2$ and the upper part of the $\beta_1$ Fermi surface where the $xs$ gap is negative. For scatterings with a momentum transfer $q_2=(\pi,0.69\pi)$ shown in Fig.~\ref{fig:fig3}b, the sign of the $s^\pm$ gap is the same on $\beta_1$ and $\beta_2$ and the vanishing of the coherence factor suppresses the resonance. However, for the extended $xs$-wave, there is a weak response, which for a slightly shifted value of momentum transfer $q\sim(\pi,0.66\pi)$ shows a resonance reflecting scattering from the bottom ``positive" gap region of the $\beta_2$ Fermi surface to the top of the $\beta_1$ Fermi surface where the $xs$ gap is negative. The Brillouin zone intensity difference plots which will be discussed below provide another way to see this.

Turning next to the $d$-wave gap, Fig.~\ref{fig:fig3}b shows a resonance for $q_2=(\pi,0.69\pi)$ associated with the scattering between the electron $\beta_1$ and $\beta_2$ Fermi surfaces. Interestingly, at a lower energy there is also a resonance for $q_2=(\pi,0.16\pi)$. This resonance at $\omega\sim\Delta_0$ corresponds to the smaller peak in
 the susceptibility seen for $n=0.09$ in Fig. \ref{fig:fig2}. 
 Here the $d$-wave gap on part of the $\alpha_2$ hole sheet is out of phase with the gap on the $\beta_1$ sheet, so the coherence factor is non-vanishing and a resonance can appear. The $xs+id$ gap also shows a resonance response for both $q_1$ and $q_2$. However, in this case, the energy of the resonance at $q_1$ is closer to that of the $xs$ gap and at $q_2$ closer to that of the $d$-wave gap as one would expect.



\begin{figure*}[ht]
\includegraphics[width=2.0\columnwidth]{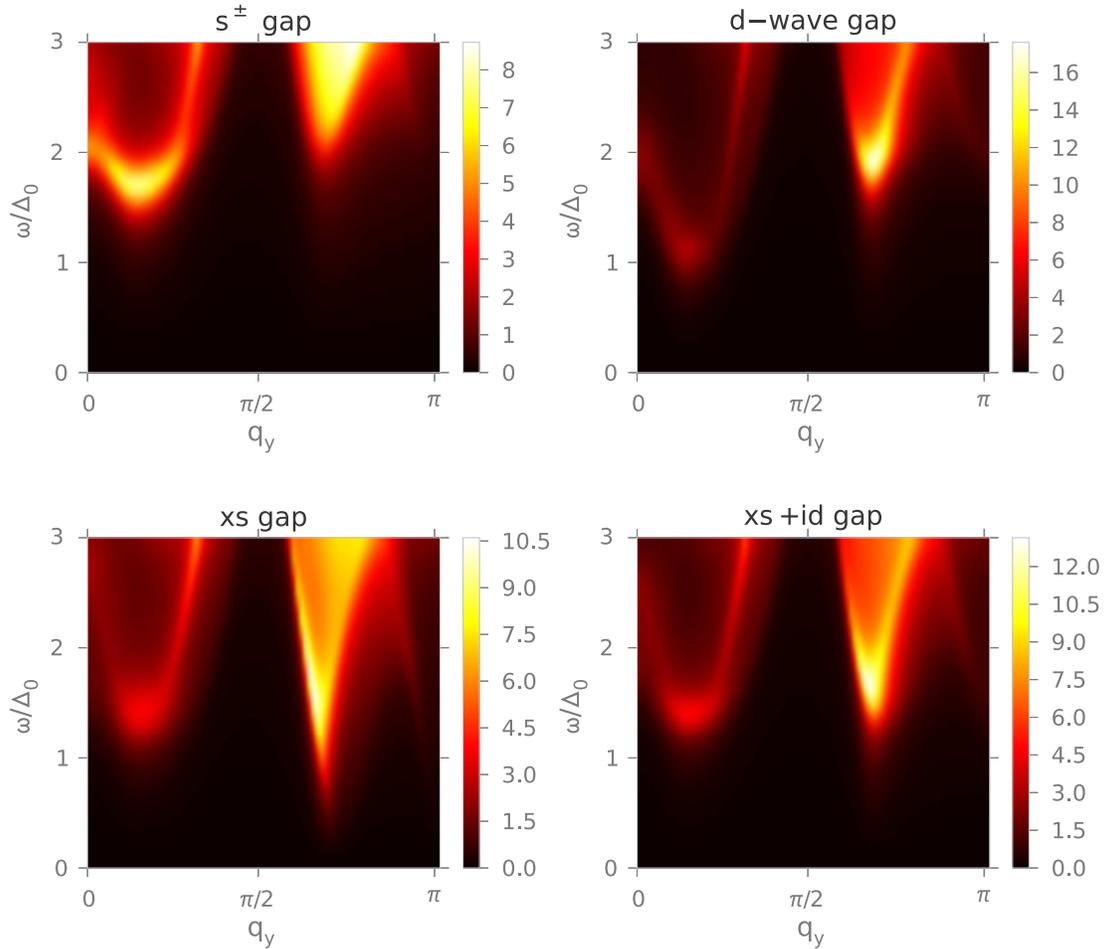}
\caption{\label{fig:fig4}
(Color online) Imaginary part of the dynamical susceptibility $\chi''_s(q,\omega)$ plotted for $q_x=\pi$
as function of $q_y,\omega$ for various model pair states for $n=0.09$.
}
\end{figure*}

Important information helping to identify the gap symmetry  and structure 
can be obtained from the qualitative way in which the resonance structures 
at the different wave vectors disperse with energy.
In Fig.~\ref{fig:fig4} we show a set of  plots of $\chi''(q,\omega)$
for the various gap structures given by Eqs.~(\ref{eqn:gaps1})--(\ref{eqn:gaps4}) along a cut with $q_x=\pi$. These
susceptibility maps
provide an alternative way of looking at the response from the fixed momentum, variable frequency plots shown in Figs.~\ref{fig:fig3}a and b. The 
map for the $s^\pm$ gap exhibits structure in the $(0,\pi)$ and $(\pi,0)$ regions associated with scattering from the $\alpha$ hole Fermi surfaces to the $\beta$ electron Fermi surfaces. In the simple isotropic $s^\pm$ case, this resonance is near $(\pi,0.16\pi)$ and $\omega\approx1.75\Delta_0$. At higher energies above $2\Delta_0$, a significant broad intensity appears also at larger values of $q_y$ due to pair breaking quasiparticle scattering processes between the $\beta$ sheets. 


The map for the extended $xs$-wave gap state exhibits a similar resonance peak to the $s^\pm$ case in the $(0,\pi)$ and $(\pi,0)$ regions due to $\alpha$ to $\beta$ scattering. However, as already seen in Fig.~\ref{fig:fig3}a, the resonance occurs at a lower energy than for the $s^\pm$ gap. In addition, the extended $xs$ state also has resonances for $q\sim(\pi,0.66\pi)$ and $(0.66\pi,\pi)$. These arise from scattering processes between the $\beta_1$ and $\beta_2$ electron sheets, whereby an electron scattered from the lower part of the $\beta_2$ Fermi surface to the upper part of the $\beta_1$ Fermi surface with $q\sim(\pi,0.66\pi)$ scatters between regions for which the $xs$ gap changes sign. Then at threshold the coherence factor for this scattering process is unity and a resonance can appear. 
In Fig. \ref{fig:fig4}, one sees that at higher energies the branch dispersing towards $q_y=\pi/2$ has larger intensity than the branch dispersing towards $q_y=\pi$.

The $d$-wave map shows a strong resonance for $\omega\lesssim 2\Delta_0$ at $(\pi,0.69\pi)$ and $(0.69\pi,\pi)$ as expected from the $\omega$ scan at $q=(\pi,0.69\pi)$ shown in Fig.~\ref{fig:fig3}b. This resonance arises from $\beta_1$ to $\beta_2$ scatterings, and is seen to have higher intensity towards $q_y=\pi$ as $\omega$ increases.  In addition, as seen in the $\omega$ scan at $q=(\pi,0.16\pi)$, the $d$-wave gap also exhibits weaker resonances associated with $\alpha_2$ to $\beta$ scattering processes. These arise because the $\pi$-phase shift in the sign of the $d$-wave gap on the $\alpha_2$ Fermi surface gives rise to a resonance for scattering an electron from the  $\alpha_2$ Fermi surface to the $\beta_1$ Fermi surface.

While the $q$-maps provide a clear way to distinguish between an $s^\pm$ and $d$-wave state, the extended $xs$ gap and the $d$-wave gap both show resonances from $\alpha$--$\beta$ and $\beta$--$\beta$ scattering.   On the basis of the differences in the spectral weight and resonance energies alone, it would be difficult to distinguish between $xs$ and $d$-wave states. However, for the $xs$ case, the $\beta$--$\beta$ resonances which onset near $(\pi,0.66\pi)$ and $(0.66\pi,\pi)$ show more intensity along the branch dispersing towards $(\pi,0)$ and $(0,\pi)$, respectively, while for the $d$-wave case the resonances, which onset near $(\pi,0.69\pi)$ and $(0.69\pi,\pi)$, are stronger towards $(\pi,\pi)$. This difference in dispersion arises from an interplay between the BCS coherence factor $(1-\frac{\Delta_\mu(k+q)\Delta_\nu(k)}{E_\mu(k+q)E_\nu(k)})$ and the energy conserving $\delta$-function $\delta(\omega-E_\mu(k+q)-E_\nu(k))$, which control the scattering phase space. For the $xs$ case, where the resonance onsets for $q\sim(\pi,0.66\pi)$, the coherence factor selects scattering processes in which electrons scatter from the bottom of the $\beta_2$ Fermi surface to the top of the $\beta_1$ Fermi surface. In this case, as $\omega$ increases, the $k_y$ separation of the scattering states decreases and the $xs$ resonance is stronger towards $(\pi,0)$. For the $d$-wave case, there is a relative change of the sign of the gap between the entire $\beta_1$ and $\beta_2$ Fermi surfaces. In this case, as $\omega$ increases and the $\k$ integration for $\chi''(q,\omega)$ runs over a range of different energy cuts of the band energies, the peak response shifts towards the commensurate wave vector $q=(\pi,\pi)$ connecting the centers of $\beta_1$ and $\beta_2$ Fermi surfaces. It is interesting to note that the recent data of Friemel et al.\cite{Inosov2} suggest a weak dispersion of the resonance towards $(\pi,\pi)$, providing additional support for  $d$-wave symmetry.

Finally, there is the question of identifying the $xs+id$ state. While the gap map for an $xs+id$ gap is clearly different from the $s^\pm$ gap and one might differentiate it from the $xs$ case by examining the dispersion of the resonances, the difficulty is separating it from the $d$-wave case. That is, while there are clearly differences between the $xs+id$ and $d$-wave gap maps shown in Fig.~\ref{fig:fig4}b, these are quantitative differences which depend upon the magnitudes of the gaps which are chosen on the various Fermi surfaces. The $xs+id$ and $d$-wave gaps can be distinguished qualitatively only at low energies, where $\chi''({\bf q},\omega)$ will reflect the full gap in the $xs+id$ case, as opposed to the nodal $d$-wave state.  For the $q$-$\omega$ maps shown in Fig.~\ref{fig:fig4}, $q_x=\pi$ and the $d$-wave nodes are not connected by any $q_y$. However, nodal excitations can be probed by neutrons at other wave vectors. In addition, other experimental probes of low-energy quasiparticles can be used to distinguish the $d$ and $xs+id$ states. 

While we have presented concrete calculations within a particular model and for a given set of interaction parameters, leading to resonances at  specific energies and wave vectors, it is important to stress that the actual resonance energies, widths, etc. may differ due to uncertainties in doping, band structure, interactions, or other details. In fact, the analysis presented here suggests that samples differing in doping by very small amounts can have different {\it symmetry} order parameters.  It is therefore  important that experiments like ARPES and neutron scattering be performed on the same sample in order to draw robust conclusions.  Here we have tried to focus on differences between various gap symmetries and structures which can allow for qualitative distinctions via measurements of the neutron resonance at different dopings, and made predictions for the evolution of this resonance with doping and energy.
In conclusion, the AFe$_2$Se$_2$ materials may exhibit an $A_{1g}$ ($s^\pm$-wave), $B_{1g}$ ($d$-wave), or possibly $xs+id$ gap. In the region where these phases are nearly degenerate, they can give rise to resonances associated with $\alpha$ to $\beta$ as well as $\beta$ to $\beta$ scattering processes. In this case, one can distinguish the $A_{1g}$ state from the $B_{1g}$ state by examining the dispersion of the $\beta$ to $\beta$ resonance.

\acknowledgements

The authors are grateful to D. Inosov and I.I. Mazin for helpful discussions.  PJH was supported by DOE DE-FG02-05ER46236. A portion of this research was conducted at the Center for Nanophase Materials Sciences, which is sponsored at Oak Ridge National Laboratory by the Scientific User Facilities Division, Office of Basic Energy Sciences, U.S. Department of Energy.

\end{document}